%% file: main.tex
\pdfoutput=1  
\documentclass[10pt,twocolumn]{article}
\usepackage[a4paper,margin=1.9cm]{geometry}
\input{preamble}
\input{preamble_sapphire}   

\title{The Documentation and Traceability Burden of India's EV Transition: A Systematisation from an Information-Systems Perspective}
\author{Dawar Jyoti Deka \and Nilesh Sarkar}
\date{3 July 2026}
\hypersetup{pdftitle={The Documentation and Traceability Burden of India's EV Transition},
            pdfauthor={Dawar Jyoti Deka and Nilesh Sarkar}}

\begin{document}
\twocolumn[{%
\begin{tcolorbox}[colback=sapphireLight, colframe=sapphireMid!55, boxrule=0.6pt,
                  arc=4pt, left=12pt, right=12pt, top=10pt, bottom=10pt, width=\textwidth,
                  before skip=0pt, after skip=6pt]
  \begin{minipage}[c]{2.05cm}\centering\includegraphics[width=1.75cm]{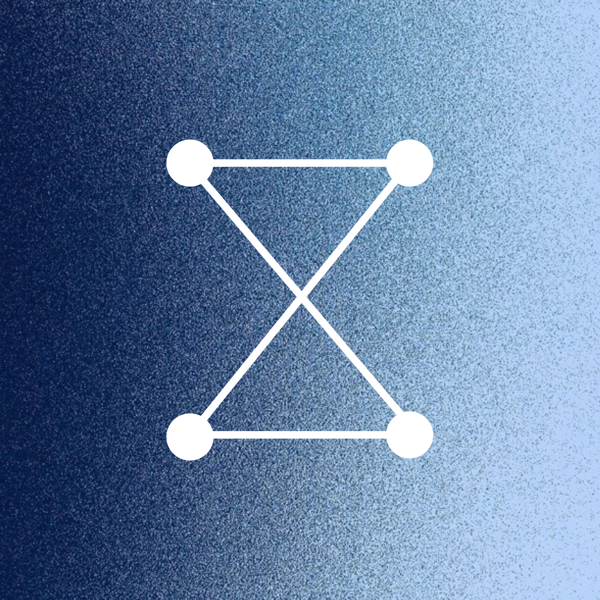}\end{minipage}%
  \hfill
  \begin{minipage}[c]{\dimexpr\linewidth-2.55cm\relax}
    \raggedright
    {\bfseries\LARGE\color{sapphireDark}The Documentation and Traceability Burden of India's EV Transition}\\[3pt]
    {\bfseries\large\color{sapphireDark}A Systematisation from an Information-Systems Perspective}
  \end{minipage}
\end{tcolorbox}
\begin{center}
  {\large Dawar Jyoti Deka \qquad Nilesh Sarkar}\\[3pt]
  {\normalsize Erd\H{o}s Systems}\\[2pt]
  {\normalsize \href{mailto:dawar@erdoslab.org}{\texttt{dawar@erdoslab.org}} \quad\textbullet\quad \href{mailto:nilesh@erdoslab.org}{\texttt{nilesh@erdoslab.org}}}\\[3pt]
  {\normalsize 3 July 2026}
\end{center}
\vspace{2pt}
}]

\begin{abstract}
\input{sections/00_abstract}
\end{abstract}

\input{sections/01_introduction}
\input{sections/02_background}

\input{sections/03_spine}
\input{tables/tab1_spine}

\input{sections/04_ev_stack}
\input{tables/tab2_obligations}

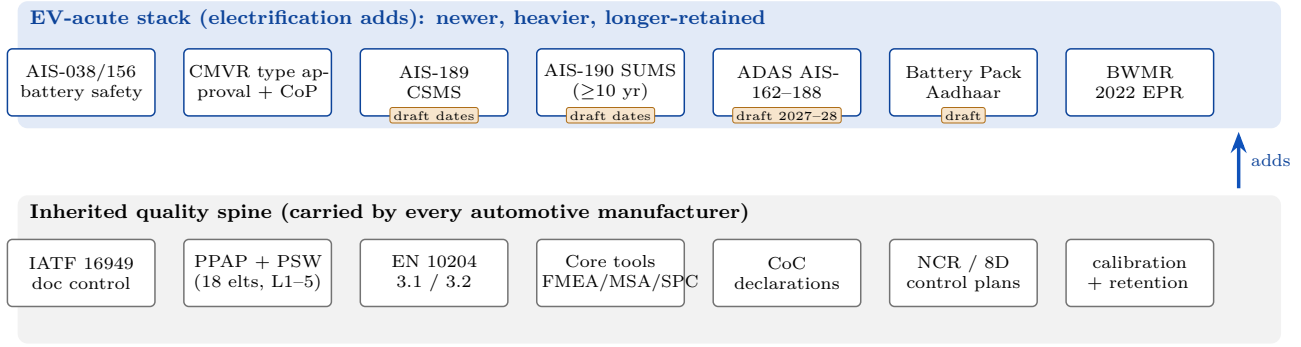
\begin{figure*}[t]
\centering
\resizebox{\textwidth}{!}{\input{figures/fig2_stack.tikz}}
\caption{The two-layer evidence burden of an Indian EV OEM. Every automotive manufacturer carries the inherited quality spine (lower band); electrification adds the EV-acute stack (upper band), newer, heavier and longer-retained. Amber tags mark instruments whose dates were draft or contested at the snapshot (\snapshotdate).}
\label{fig:stack}
\end{figure*}

\input{sections/05_failures}
\input{tables/tab3_cases}

\begin{figure*}[t]
\centering
\resizebox{\textwidth}{!}{\input{figures/fig3_timeline.tikz}}
\caption{Failure events (above the axis) and regulatory response (below), 2022 to 2026. The FAME-II thread shows one evidence base yielding divergent state outcomes over three years; case letters key to \cref{fig:lifecycle} and \cref{tab:cases}.}
\label{fig:timeline}
\end{figure*}

\begin{figure*}[t]
\centering
\resizebox{\textwidth}{!}{\input{figures/fig1_lifecycle.tikz}}
\caption{The compliance-document lifecycle. Six stages with a Change/Revalidate return loop; M marks stages still manual today; the cut on the Approve-to-Record edge marks the approval-to-artefact decoupling; amber pins are the failure loci A--D of \cref{sec:failures}.}
\label{fig:lifecycle}
\end{figure*}
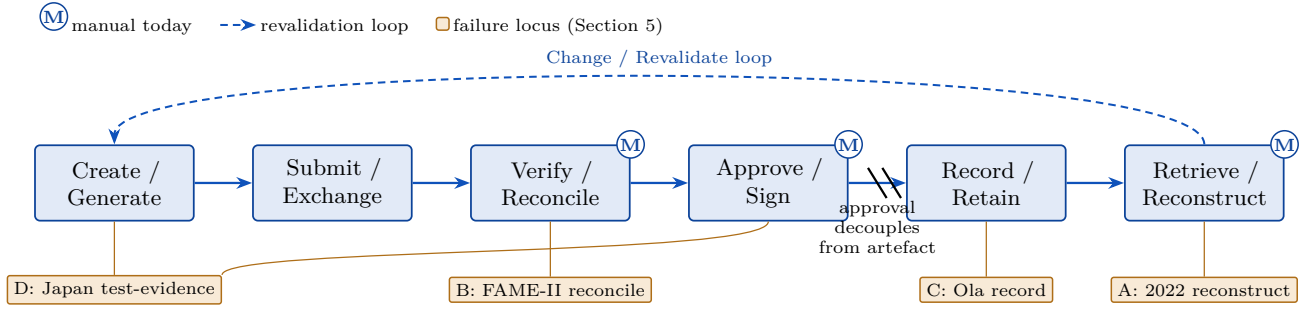

\input{sections/06_lifecycle}
\input{sections/07_lens}
\input{sections/08_requirements}
\input{sections/09_open_problems}
\input{sections/10_conclusion}

\section*{Competing interests}
The authors are affiliated with Erd\H{o}s Systems, a company developing software for
compliance-document processing in the automotive sector; the first author is its
founder. This work received no external funding.

\bibliographystyle{plainnat}
\bibliography{refs}

\onecolumn
\appendix
\input{sections/appendix_a_standards}
\input{sections/appendix_b_taxonomy}
\input{sections/appendix_c_chronologies}
\input{sections/appendix_d_littable}

\end{document}

%% file: preamble.tex
\usepackage[T1]{fontenc}
\usepackage[utf8]{inputenc}
\usepackage{lmodern}
\usepackage{microtype}
\usepackage{booktabs}
\usepackage{array}
\usepackage{tabularx}
\usepackage{longtable}
\usepackage{multirow}
\usepackage{ragged2e}
\usepackage{amsmath}
\usepackage{amssymb}
\usepackage{graphicx}
\usepackage{xcolor}
\usepackage{tikz}
\usetikzlibrary{arrows.meta,positioning,shapes.geometric,calc,fit,backgrounds}
\usepackage[numbers,sort&compress]{natbib}
\usepackage[colorlinks=true,allcolors=blue!55!black,breaklinks=true]{hyperref}
\usepackage{cleveref}
\usepackage{enumitem}
\setlist{nosep,leftmargin=1.2em}

\usepackage{balance}

\newcolumntype{L}[1]{>{\RaggedRight\arraybackslash}p{#1}}

\crefname{table}{Table}{Tables}
\crefname{figure}{Figure}{Figures}
\crefname{section}{Section}{Sections}

\newcommand{\snapshotdate}{3 July 2026}

%% file: preamble_sapphire.tex
\ifdefined\sapphireloaded \fi
\def\sapphireloaded{}

\definecolor{sapphire}{HTML}{0F52BA}      
\colorlet{sapphireDark}{sapphire!82!black}
\colorlet{sapphireMid}{sapphire}
\colorlet{sapphireLight}{sapphire!12}
\definecolor{amber}{HTML}{D98A1E}          

\usepackage{colortbl}      
\usepackage{sectsty}       
\usepackage{caption}       
\usepackage{xurl}          
\usepackage{tcolorbox}     

\sectionfont{\color{sapphireDark}}
\subsectionfont{\color{sapphireDark}}
\paragraphfont{\color{sapphireDark}}

\hypersetup{linkcolor=sapphireDark, citecolor=sapphireDark, urlcolor=sapphireMid,
            filecolor=sapphireDark}

\providecommand{\practitionermode}{0}

\ifnum\practitionermode=1
  \newcommand{\caseinsight}[1]{\par\smallskip\noindent
    \fcolorbox{sapphireMid}{sapphireLight}{%
      \parbox{\dimexpr\columnwidth-2\fboxsep-2\fboxrule}{%
        \small\textbf{\textcolor{sapphireDark}{Insight.}}\ #1}}\par\smallskip}
  \usepackage{fancyhdr}
\else
  \newcommand{\caseinsight}[1]{#1}
\fi

\newcommand{\headrow}{\rowcolor{sapphireLight}}

%% file: sections/00_abstract.tex
India's electric-vehicle transition is usually told through batteries, cost and charging. Less visibly, it is a story about documentation. Electrification layers a new and fast-growing burden of compliance evidence, type-approval test reports, battery-safety and thermal-propagation records, cyber-security and software-update documentation, and supplier-quality packages, on top of the evidence apparatus the automotive industry already carries, with retention horizons that now reach a decade past end of production. This burden is studied from the testing side but has never been systematised as an information-systems problem, so its verification bottlenecks, architecture requirements and open questions remain unmapped. We close that gap for the Indian EV OEM with four contributions: a two-layer systematisation of evidence obligations organised by artefact, producer, verifier, trigger and retention; a lifecycle model of the compliance document with manual bottlenecks and four observed failure loci marked; an analytic lens, documentation exergy destruction, that classifies document families and explains systematic under-resourcing; and a research agenda spanning reconciliation accuracy, human-in-the-loop reliability, cross-party provenance, long-horizon integrity, burden measurement and machine-verifiable regulation. Three Indian failure cases and one Japanese ground the analysis.

%% file: sections/01_introduction.tex
\section{Introduction}
\label{sec:intro}

In the spring of 2022 a series of electric two-wheelers caught fire in India. Within weeks, three makers had recalled 6,656 scooters between them, Okinawa Autotech 3,215, Pure EV 2,000 and Ola Electric 1,441, and the government had issued show-cause notices warning of penal action \citep{recalls2022,evfires_committee2022}. A regulator investigating such an incident asks a sequence of ordinary questions. Which cell batch was in the vehicle that burned? Which test report certified that batch? Which supplier produced it, under which version of the battery-management firmware, and who approved that version? Where is the record? Each question is a request to retrieve a specific artefact and reconcile it against others, across the boundaries of a cell maker, a pack assembler, an OEM, a test agency and a regulator. When the artefacts are scattered across systems and the approvals live apart from the things they approve, answering takes weeks, and sometimes the answer cannot be reconstructed at all.

The electric-vehicle transition is usually told as a story about batteries, cost and charging infrastructure. It is also, and far less visibly, a story about documentation. Electrification does not remove the automotive industry's inherited evidence burden; it adds a new and acute layer on top of it. To the supplier-quality spine that any car maker already carries, an EV OEM must add traction-battery safety and thermal-propagation evidence, cyber-security and software-update records with retention horizons measured in decades, and a widening set of driver-assistance mandates. The result is a documentation and traceability burden that is new, heavy and growing, and that is currently studied only from the side of what the tests require, not from the side of how the evidence is produced, verified, approved, retained and retrieved.

That this pattern is neither India-specific nor unique to electric vehicles is shown by Japan: on 3 June 2024 Toyota apologised for certification irregularities affecting seven models and suspended production of three, one of several Japanese makers found that year to have submitted irregular certification data, a failure of the evidence procedure rather than of the engineering \citep{cnbc_toyota2024,fortune_toyota2024,cnn_toyota2024}.

Four adjacent literatures touch this space and each misses its centre. Regulatory and transport-policy research covers what the standards test, not the document lifecycle they impose. Battery-safety and automotive-engineering research treats documentation as a by-product. The information-systems literatures of business-process compliance, RegTech and intelligent document processing are generic, with nothing specific to the automotive evidence chain. Records-management and archival science study record-keeping in general but have not mapped this domain's artefacts, obligations or failures. The intersection, an information-systems systematisation of the automotive and EV compliance-document lifecycle, is empty.

This paper fills it, for the Indian EV OEM. We make four contributions. \textbf{C1}: a two-layer systematisation of the evidence obligations, the inherited quality spine and the EV-acute regulatory stack, organised by artefact, producer, verifier, trigger and retention horizon. \textbf{C2}: a lifecycle model of the compliance document, with the points where work is manual today, the point where approval decouples from the artefact, and four observed failure loci marked. \textbf{C3}: an analytic lens, documentation exergy destruction, that classifies document families by the mandatory verification and record-keeping work they carry and explains why that work is systematically under-resourced. \textbf{C4}: an open-problems agenda spanning reconciliation accuracy, human-in-the-loop verification reliability, cross-party approval provenance, long-horizon evidence integrity, the empirical measurement of documentation burden, and machine-verifiable regulatory design.

The remainder proceeds as follows. Section 2 sets scope and the related-work gap. Section 3 compresses the inherited quality spine and Section 4 systematises the EV-acute stack. Section 5 presents four failure cases; Section 6 abstracts them into the lifecycle model; Section 7 develops the lens. Section 8 derives evidence-architecture requirements, Section 9 the research agenda. This is the first of a planned pair; a companion paper develops the general-automotive quality spine and a global failure-mode taxonomy.

%% file: sections/02_background.tex
\section{Background, scope and related work}
\label{sec:background}

\textbf{Scope.} We take the perspective of an original equipment manufacturer building electric vehicles and new-mobility products in India, and we study one object: the compliance and supplier-quality document domain, the artefacts that must be produced, exchanged, verified, approved, retained and retrieved for a vehicle to be legitimately built and sold. We do not study battery chemistry, cost or charging, except where they generate documents. The regulatory position is stated as of the snapshot date; Section 4 fixes it precisely. India is the lead context; the European battery passport enters only as one comparator.

\textbf{Four literatures, and the gap.} The target space is the automotive and EV compliance-document lifecycle, its evidence artefacts and obligations, and its observed failures, seen through an information-systems lens. Four literatures crowd around it and each holds one axis while dropping the others. A full nearest-neighbour table of the works surveyed appears in Appendix D; we summarise the argument here.

Regulatory, certification and transport-policy research comes closest on domain but stops at the standards themselves. It establishes what tests and criteria the regulations demand and how frameworks interlock, for example comparing UN Regulation 155 with ISO/SAE 21434 or deriving type-approval criteria for cyber-security and software-update management, or it reviews India's EV policy at the macro level. Compliance evidence appears only as a by-product of type approval, never as a managed artefact chain. The single recent work that names certification evidence explicitly is confined to one regulation and framed as an engineering process model, not a systematisation of the document lifecycle.

Battery-safety and automotive-engineering research is a physics-and-test-method literature: abuse and thermal-runaway testing, battery-management-system estimation, functional-safety work products under ISO 26262, and reviews of Indian battery-safety standards. Across all of it, the test reports, safety files and work products these activities generate are treated as an afterthought rather than as the object of study.

The information-systems literature supplies the lens but never this object. Business-process compliance formalises the alignment of process execution with normative specifications \citep{sadiq2007,hashmi2018}. RegTech applies technology to regulatory monitoring and reporting, oriented mainly to financial services \citep{arner2017}. Intelligent document processing extracts fields from generic forms with no compliance semantics \citep{subramani2020}. Each treats compliance, documents or traceability generically; none takes the automotive or EV compliance-evidence artefact, its obligations and its failures as the unit of analysis. Records-management and archival science supply the recordkeeping and authenticity vocabulary but are, by design, domain-agnostic \citep{iso15489,duranti1995}.

The intersection is therefore empty. No work occupies the centre where a domain-specific artefact chain, an information-systems and records lens, and an India-led compliance-document-lifecycle framing meet. We searched for one; if it existed, this paper would be an extension rather than a systematisation.

\textbf{Two nearest-neighbour objections.} A competent reviewer raises two immediately.

*Is this just records management?* No. ISO 15489 sets out concepts and principles for creating and managing authentic, reliable records, and the InterPARES programme and Duranti develop the theory of authenticity and trust for records maintained in digital form \citep{iso15489,duranti1995}. But these frameworks are deliberately domain-agnostic: ISO 15489 applies to records in any format or technological environment, and neither it nor InterPARES enumerates this domain's specific artefacts (homologation files, conformity-of-production records, AIS and CMVR test reports, cyber-security and software-update evidence, PPAP packages), the cross-party evidence obligations that produce and retain them, or the documentation failures observed in India's EV transition. This paper is an instantiation, not a restatement: it does the domain-specific work the general frameworks leave undone.

*Is this just business-process compliance?* No. Business-process compliance asks whether a process conforms to a norm; it models control objectives and checks execution against a specification \citep{sadiq2007,hashmi2018}. It is general and domain-independent, and it never treats the compliance-evidence document as the unit of analysis. It does not model the automotive evidence chain, the cross-party artefacts it produces (OEM, tier suppliers, test houses and certification authorities each hold and hand off distinct records), the obligations attached to those artefacts, or the documented failures where the chain breaks. We ask a different question: not does the process conform, but what evidence must exist, who produces, holds and retains it, and where does the evidence chain fail.

%% file: sections/03_spine.tex
\section{The inherited quality-document spine}
\label{sec:spine}

An EV OEM inherits the automotive industry's supplier-quality apparatus in full, because it still sources motors, cells, battery-management boards and power electronics from a tiered supply base. We compress it here; the companion paper develops it in depth, and Table 1 summarises the artefacts.

The governing standard, IATF 16949:2016, imposes documented-information control, a defined record-retention policy, and identification and traceability \citep{iatf16949}. Within it, the Production Part Approval Process (PPAP) is the central evidence package: up to eighteen elements defined by the AIAG manual \citep{aiagppap}, submitted at one of five levels that set how much is sent versus retained, with Level 3 the common default \citep{aiagppap}. The Part Submission Warrant (PSW) is the summary sign-off, and its disposition gates shipment \citep{aiagppap}. A PPAP is triggered by a new part, a design or process change, or a change of supplier or material source \citep{aiagppap}, so the package is regenerated across a part's life, not produced once.

Around PPAP sit the conformance artefacts: a Certificate of Conformity, the supplier's declaration that a part meets requirements \citep{isoiec17050}; EN 10204 inspection documents, type 3.1 from the manufacturer's independent representative and 3.2 with third-party validation \citep{en10204}; and the core-tools records any auditor expects, nonconformance reports, 8D, FMEA and PFMEA, control plans, MSA and Gage R\&R, SPC and calibration \citep{iatf16949,aiagppap}. These are generated and reconciled across the enterprise's MES, ERP, PLM and QMS systems \citep{qmsintegration}, the first appearance of a theme this paper returns to: the evidence is distributed, and its coherence is a cross-system property.

The cost of this spine is concrete, and much of it is machine-checkable. A Level 3 PPAP takes weeks, roughly two to three for a simple part and eight to twelve or more for a complex one, and no standard fixes the figure \citep{ppapcycle}. Most of that time is verification. The rejection patterns are objective: a missing or invalid signature, a wrong drawing revision, a PSW inconsistency, incomplete dimensional coverage, a capability index below the conventional 1.67, with 1.33 for significant characteristics \citep{ppapreject,aiagppap}. A reviewer rejecting a package for a revision mismatch is performing exactly the reconciliation that recurs, at higher stakes, throughout the EV-acute stack.

Everything above is inherited the moment an EV OEM sources a motor, a cell or a power-electronics module, so findings generalise across automotive manufacturing. Electrification does not replace this spine; it adds a heavy new layer, which the next section systematises.

%% file: tables/tab1_spine.tex
\begin{table}[t]
\centering
\footnotesize
\setlength{\tabcolsep}{4pt}
\renewcommand{\arraystretch}{1.15}
\caption{The inherited automotive quality-document spine an EV OEM carries over from conventional automotive supply. Governing references are the AIAG core-tools manuals and IATF~16949; the depth is developed in the companion paper.}
\label{tab:spine}
\begin{tabularx}{\columnwidth}{@{}L{2.15cm} L{1.75cm} X L{1.7cm}@{}}
\toprule
\headrow
\textbf{Artefact} & \textbf{Governing ref.} & \textbf{Purpose / trigger} & \textbf{Typical verifier} \\
\midrule
PPAP package (up to 18 elements, 5 levels) & AIAG PPAP & Proves a process is capable and ready; triggered by a new part or a design, process or supplier change & Customer / OEM SQE \\
Part Submission Warrant (PSW) & AIAG PPAP & Summary sign-off; its disposition gates shipment & Customer / OEM SQE \\
Certificate of Conformity (CoC) & ISO/IEC 17050 & Supplier declaration that a part meets requirements & Receiving quality \\
Material certificate 3.1 / 3.2 & EN 10204 & Inspection evidence; 3.2 adds independent sign-off & OEM / third party \\
FMEA / PFMEA, control plan & AIAG core tools & Risk analysis feeding inspection controls & OEM / auditor \\
MSA / Gage R\&R, SPC (Cpk) & AIAG core tools & Measurement-system and capability evidence ($\geq$1.67 for special characteristics) & OEM / auditor \\
NCR, 8D, calibration records & IATF 16949 & Nonconformance, problem-solving and traceability records & OEM / auditor \\
Retention and traceability policy & IATF 16949 & Controls documented information, retention and identification & Certification body \\
\bottomrule
\end{tabularx}
\end{table}

%% file: sections/04_ev_stack.tex
\section{The EV-acute evidence stack}
\label{sec:evstack}

Regulatory position stated as of the snapshot date, 3 July 2026. The regulatory landscape moves; a delta sweep for the six months preceding this date found only draft advances, timeline reaffirmations and subsidy extensions, none of which reverses a position stated here \citep{deltasweep2026}. We organise the stack by evidence obligation rather than by standard, because the same artefact often serves several instruments and the same instrument spawns several artefacts. Table 2 is the reference and Figure~\ref{fig:stack} sets the two layers side by side; the prose gives the obligations and their provenance.

\textbf{Traction-battery safety evidence.} Two standards anchor the stack. AIS-038 (Rev.2) sets construction and functional-safety requirements for the traction battery and electric power train of M and N category vehicles \citep{ais038}; AIS-156 does the same for the L category, the two-wheelers, three-wheelers and quadricycles that dominate India's EV volume \citep{ais156}. Following the 2022 fires, a MoRTH amendment notified in September 2022 added battery-safety requirements on a phased mandatory basis, Phase 1 from 1 December 2022 and Phase 2 from 31 March 2023 \citep{morth_amend2022}, and the standards continued to be amended thereafter, AIS-156 through Amendment 4 in December 2023 \citep{ais156}. The evidence these create is substantial. A thermal-propagation test must show that a single-cell thermal runaway does not propagate to fire or explosion through the pack, accompanied by an audio-visual warning system to alert occupants early \citep{ais156}. The pack must meet IPX7 water-ingress protection, tested at full state of charge \citep{ais156}. The producer must maintain traceability at the pack, cell, BMS and charger level; notably, the earlier explicit requirement for an RFID tag was removed in the September 2022 amendment and replaced by this general traceability obligation \citep{morth_amend2022}, a reminder that the evidence requirements themselves are a moving target. A smart, microprocessor-based BMS with full cell-level protections, at least four in-pack temperature sensors and its own EMC evidence is required \citep{ais156}.

\textbf{Type approval and conformity of production.} These artefacts do not float free; they are consumed by the homologation machinery. Battery-safety compliance is enforced through CMVR type approval: prototype testing is performed by notified agencies, ARAI and ICAT, under Rule 126 and AIS-049 (Rev.1), and MoRTH extended conformity-of-production obligations to traction batteries by amending Rule 124 of the CMVR through notification GSR 659(E), effective 1 October 2022 \citep{cmvr_gsr659e}. Type approval produces a certificate that must be reconciled against the type it certifies; conformity of production requires continuing evidence that serial output matches the approved prototype. The producer is thus on the hook for an initial evidence pack and a continuing one.

\textbf{Cyber-security and software-update evidence.} Software adds two instruments modelled on United Nations regulations. AIS-189 is India's Cyber Security and Cyber Security Management System standard, derived from UNECE Regulation 155 \citep{ais189}; AIS-190 is the Software Update and Software Update Management System standard, derived from UNECE Regulation 156 \citep{ais190}. Their effective dates are not yet fixed law: no MoRTH gazette had settled them at the snapshot, and reported schedules conflict, with an industry-proposed timeline phasing in OTA-capable new models from about October 2027 and later cohorts through 2030, while some secondary sources cite October 2025 and 2028 \citep{ais189,ais190}. What is fixed, and load-bearing for the research agenda, is a retention obligation. Under AIS-190, mirroring UN R156, the type-approval documentation package covering the SUMS and the software version and update records must be retained and remain available for at least ten years, counted not from creation but from the time production of the vehicle type is definitely discontinued, and the obligation binds both the test agency and the manufacturer \citep{ais190}. An OEM must therefore keep a software-provenance record intact and retrievable for a horizon that can exceed the working life of the systems that created it.

\textbf{Driver-assistance mandates.} A widening ADAS suite adds function-level evidence. Under draft notification GSR 184(E) of 20 March 2025, India is phasing in AIS-162 (advanced emergency braking), AIS-184 (driver drowsiness and attention warning), AIS-186 (blind-spot information, based on UN R151), AIS-187 (moving-off information, based on UN R159) and AIS-188 (lane-departure warning) for M2, M3, N2 and N3 vehicles \citep{morth_gsr184e}. The phase-in has itself shifted: against the draft's 2026 dates, a February 2026 ministry reply placed braking obligations from 1 October 2027 and the warning functions from 1 January 2028 \citep{deltasweep2026}. Each function adds a type-approval evidence trail to the pack.

\textbf{Traceability, passports and producer responsibility.} Two further obligations govern the battery over its life. India's battery-passport framework exists, as of the snapshot, only as a draft. The "Battery Pack Aadhaar" guidelines, developed by a MoRTH committee under the Office of the Principal Scientific Adviser's e-mobility roadmap and proposed for the AIS route, would assign each pack a unique identity and carry lifecycle traceability data \citep{bpan_draft}. It remains under consultation, not a notified rule, with a reported 2027 target \citep{bpan_draft}. Separately and already in force, the Battery Waste Management Rules 2022 impose Extended Producer Responsibility and require producers, recyclers and refurbishers to register with the CPCB and maintain collection and recycling records \citep{bwmr2022}. For comparison, the European Union's Batteries Regulation 2023/1542 requires a battery passport from 18 February 2027 for electric-vehicle, light-means-of-transport and larger industrial batteries \citep{eu2023_1542}, and the Global Battery Alliance has run passport pilots since January 2023 \citep{gba2023}; India's trajectory is toward, but has not reached, a comparable mandate.

\textbf{A note on the subsidy layer.} The demand-side schemes that shaped supplier behaviour are relevant because their localisation conditions themselves generated evidence obligations, as Section 5 shows. FAME-II ended on 31 March 2024 and was succeeded by the interim EMPS 2024 and then PM E-DRIVE, a Rs 10,900 crore scheme effective from October 2024 \citep{pmedrive}.

Taken together, the stack is not a list of tests but a set of standing obligations to produce, verify, approve, retain and retrieve specific artefacts, across parties, on horizons up to a decade past end of production. The next sections show what happens when that obligation set is managed by hand.

%% file: tables/tab2_obligations.tex
\begin{table*}[t]
\centering
\footnotesize
\setlength{\tabcolsep}{4pt}
\renewcommand{\arraystretch}{1.15}
\caption{The EV-acute evidence stack for an Indian EV OEM, organised by evidence obligation rather than by standard. Regulatory position stated as of the snapshot date (\snapshotdate). Every entry is traceable to the claims ledger; the supporting citations are given in the running text of \cref{sec:evstack}.}
\label{tab:evstack}
\begin{tabularx}{\textwidth}{@{}L{2.35cm} L{1.75cm} L{3.5cm} L{2.5cm} L{2.2cm} X@{}}
\toprule
\headrow
\textbf{Instrument} & \textbf{Scope} & \textbf{Evidence artefact(s) created} & \textbf{Producer $\rightarrow$ verifier} & \textbf{Trigger} & \textbf{Retention / effective date} \\
\midrule
AIS-038 (Rev.2) traction-battery safety & M, N (car, bus, truck) EVs & Type-approval test report; thermal-propagation and IPX7 test evidence; pack/cell/BMS/charger traceability records & OEM/supplier $\rightarrow$ test agency (ARAI, ICAT), MoRTH & New type; design change & Type-approval validity; phased mandatory from 1 Dec 2022 (Phase 1) and 31 Mar 2023 (Phase 2) \\
AIS-156 REESS and power-train safety & L (2-, 3-wheelers, quadricycles) & As above, plus audio-visual thermal-warning evidence and BMS/temperature-sensor records & OEM/supplier $\rightarrow$ test agency (ARAI, ICAT), MoRTH & New type; design change & As above; amended through Amendment 4 (Dec 2023) \\
CMVR type approval and conformity of production & All categories & Type-approval certificate (Rule 126, AIS-049 Rev.1); conformity-of-production records & OEM/supplier $\rightarrow$ test agency, MoRTH & Type approval; serial production & CoP for traction batteries mandated via GSR 659(E), from 1 Oct 2022 \\
AIS-189 cyber security (CSMS) & Applicable road vehicles & CSMS process evidence; risk-assessment and management records (modelled on UN R155) & OEM $\rightarrow$ test agency, MoRTH & Type approval; new cyber risk & Proposed phased schedule; final gazette dates not settled (see note) \\
AIS-190 software update (SUMS) & Applicable road vehicles & SUMS documentation package; software version and update-integrity records (modelled on UN R156) & OEM $\rightarrow$ test agency, MoRTH & Type approval; each software update & Documentation retained $\geq$10 years after production of the type is discontinued; dates proposed, not settled \\
ADAS suite AIS-162/184/186/187/188 & M2, M3, N2, N3 (buses, trucks) & Function type-approval evidence for AEBS, drowsiness, blind-spot, moving-off, lane-departure warning & OEM/supplier $\rightarrow$ test agency & New / existing models & Draft GSR 184(E) (20 Mar 2025); phase-in reaffirmed for 2027--2028 (see note) \\
Battery Pack Aadhaar (draft) & L, M, N EV and $>$2 kWh industrial batteries & Unique battery identity (21-character code plus QR); lifecycle traceability data & OEM/producer $\rightarrow$ MoRTH, AIS committee & Battery placed on market & DRAFT under consultation (OM 17 Sep 2025); reported 2027 target \\
Battery Waste Management Rules 2022 (EPR) & Battery producers & EPR registration; collection, recycling and refurbishment records & Producer/recycler $\rightarrow$ CPCB & Placing batteries on market & CPCB registration; in force \\
\bottomrule
\end{tabularx}

\par\smallskip\noindent{\footnotesize\emph{Note.} AIS-189/AIS-190 and the ADAS suite have contested effective dates: no MoRTH gazette had fixed the CSMS/SUMS dates at the snapshot; an industry-proposed schedule phases OTA-capable new models from about October 2027, and a February 2026 ministry reply placed ADAS braking obligations from 1 October 2027 and warning obligations from 1 January 2028. The EU Batteries Regulation (2023/1542) requires a battery passport from 18 February 2027 for EV, LMT and $>$2 kWh industrial batteries, offered here only as a comparator.}
\end{table*}

%% file: figures/fig2_stack.tikz
\begin{tikzpicture}[
  font=\footnotesize,
  spec/.style={rectangle, rounded corners=2pt, draw=sapphireDark, line width=0.6pt,
               fill=white, minimum height=0.95cm, text width=1.95cm, align=center,
               inner sep=2pt, font=\scriptsize},
  base/.style={spec, draw=black!55},
  tag/.style={rectangle, rounded corners=1pt, fill=amber!22, draw=amber!80!black,
              line width=0.4pt, inner sep=1.4pt, font=\tiny, text=black},
]
\def\xa{0} \def\xb{2.5} \def\xc{5.0} \def\xd{7.5} \def\xe{10.0} \def\xf{12.5} \def\xg{15.0}

\begin{scope}[on background layer]
  \fill[sapphireLight, rounded corners=4pt] (-0.9,2.35) rectangle (17.0,4.15);
  \fill[black!5, rounded corners=4pt]        (-0.9,-0.7) rectangle (17.0,1.4);
\end{scope}
\node[anchor=west, text=sapphireDark, font=\footnotesize\bfseries] at (-0.85,3.9)
  {EV-acute stack (electrification adds): newer, heavier, longer-retained};
\node[anchor=west, font=\footnotesize\bfseries] at (-0.85,1.15)
  {Inherited quality spine (carried by every automotive manufacturer)};

\node[spec] (a1) at (\xa,3.0) {AIS-038/156\\ battery safety};
\node[spec] (a2) at (\xb,3.0) {CMVR type approval + CoP};
\node[spec] (a3) at (\xc,3.0) {AIS-189\\ CSMS};
\node[spec] (a4) at (\xd,3.0) {AIS-190 SUMS\\ ($\geq$10 yr)};
\node[spec] (a5) at (\xe,3.0) {ADAS AIS-162--188};
\node[spec] (a6) at (\xf,3.0) {Battery Pack Aadhaar};
\node[spec] (a7) at (\xg,3.0) {BWMR 2022 EPR};
\node[tag] at ($(a3.south)+(0,0.02)$) {draft dates};
\node[tag] at ($(a4.south)+(0,0.02)$) {draft dates};
\node[tag] at ($(a5.south)+(0,0.02)$) {draft 2027--28};
\node[tag] at ($(a6.south)+(0,0.02)$) {draft};

\node[base] (b1) at (\xa,0.3) {IATF 16949\\ doc control};
\node[base] (b2) at (\xb,0.3) {PPAP + PSW\\ (18 elts, L1--5)};
\node[base] (b3) at (\xc,0.3) {EN 10204\\ 3.1 / 3.2};
\node[base] (b4) at (\xd,0.3) {Core tools\\ FMEA/MSA/SPC};
\node[base] (b5) at (\xe,0.3) {CoC\\ declarations};
\node[base] (b6) at (\xf,0.3) {NCR / 8D\\ control plans};
\node[base] (b7) at (\xg,0.3) {calibration + retention};

\draw[-{Stealth[length=3mm]}, sapphireMid, line width=1.4pt] (16.4,1.5) -- (16.4,2.3)
  node[midway, right=1pt, font=\scriptsize, text=sapphireDark, align=left] {adds};

\end{tikzpicture}

%% file: sections/05_failures.tex
\section{Failure cases}
\label{sec:failures}

We present four documented cases. Each is drawn from the public record with cited sources; where a matter is contested, we state it as contested and give the company's position; reported investigations are described as reported, not as findings. Each case closes by pinning the failure to a stage of the lifecycle model of Section 6. Table 3 maps the set, and Figure~\ref{fig:timeline} places the cases and the regulatory response on one timeline.

\textbf{Case A: the 2022 fires and recalls.} In April 2022, following a cluster of fire incidents, three makers recalled electric two-wheelers: Okinawa Autotech 3,215 units on 16 April, Pure EV 2,000 on 21 April, and Ola Electric 1,441 on 23 April, a total the government put at 6,656 \citep{recalls2022}. The government issued show-cause notices warning of penal action \citep{recalls2022,evfires_committee2022} and convened an expert committee, including DRDO scientists alongside academics from IISc Bangalore, ARCI Hyderabad and IIT-Madras, to probe the fires \citep{evfires_committee2022}; the AIS-156 and AIS-038 amendments of Section 4 followed as the regulatory response \citep{morth_amend2022}. The information-systems point is what the investigation required. To answer which cell batch, in which vehicle, certified by which test report, from which supplier, under which BMS firmware version, approved by whom, the evidence chain had to be reconstructed after the fact and across the systems of several parties. \caseinsight{That is the lifecycle model's final stage, Retrieve and Reconstruct, failing under load.}

\textbf{Case B: the FAME-II localisation dispute.} Here the sequence is the finding. FAME-II was a central subsidy scheme with an outlay of about Rs 10,000 crore from 2019, later enhanced \citep{bt_recovery2023}. The government probed around a dozen electric two-wheeler makers over allegedly falsified localisation claims under the scheme's phased-manufacturing conditions \citep{bt_recovery2023}, and subsidy payments were put on hold for several makers, including Hero Electric and Okinawa, during 2022 to 2023 \citep{bs_famedenial2023}. In May 2023 the Ministry of Heavy Industries was reported to have issued show-cause notices to six makers seeking refund of about Rs 469 crore in combined subsidy, and three of them later returned the subsidy with interest \citep{bt_recovery2023}. Recovery was pursued against Hero Electric, Okinawa Autotech and Benling India, with the demand figures reported inconsistently across outlets and dates \citep{bt_recovery2023,inc42_cleanchit}. The companies denied violating the localisation requirements, and testing-agency reviews were cited as finding their parts locally sourced \citep{bs_famedenial2023}. The state's own conclusions then diverged. In 2024 the ministry deregistered Hero Electric, Okinawa and Benling and issued debarment orders dated 27 March 2024 against Hero Electric and Benling \citep{inc42_cleanchit}. Yet in April 2024 a ministry committee gave Hero Electric and Okinawa a clean chit, citing ARAI's inspection report of 13 May 2022 and the joint ARAI and ICAT report of 8 August 2022 \citep{bt_cleanchit2024,inc42_cleanchit}. The committee found that the scheme's notifications and phased-manufacturing guidelines lacked elaboration on localisation, causing confusion among stakeholders including the ministry itself \citep{bt_cleanchit2024,inc42_cleanchit}. A Serious Fraud Investigation Office inquiry was subsequently reported, with search operations at the three firms reported from early December 2024 over about Rs 297 crore of allegedly wrongly availed subsidy \citep{sfio_probe}. Over three years, the same evidence base produced a debarment, a clean chit and a reported fraud inquiry. \caseinsight{When localisation evidence architecture is weak enough that even the referee's agencies cannot settle the question, the failure is at Verify and Reconcile.}

\textbf{Case C: the Ola February 2025 registration gap.} Ola Electric announced about 25,000 units for February 2025, while the government's VAHAN registry recorded roughly 8,600 registrations for the month \citep{bs_ola_feb2025}. The company's stated explanation is central and we give it first: in a 19 February 2025 exchange filing it attributed the gap to a temporary registration backlog caused by renegotiating contracts with its registration vendors, Rosmerta Digital Services and Shimnit India, said registrations would be temporarily affected while sales remained strong, and framed the announced figure as bookings \citep{ola_filing2025}. Reporting indicated the figure included bookings for not-yet-delivered models, about 10,866 third-generation scooters and 1,395 Roadster X motorcycles, after which the government asked the company, in a 31 March 2025 letter, to restate February figures on an invoiced basis \citep{outlook_ola2025}. In the same weeks, RTO inspections in Maharashtra were reported to have seized scooters over trade-certificate and documentation issues, an initial 36 across Mumbai and Pune escalating to about 192 by mid-April \citep{bs_ola_feb2025}. The company's FY25 annual report later recorded a statutory-auditor finding of a material weakness in physical-verification controls at a subsidiary, covering about Rs 362 crore of vehicles and parts not physically verified \citep{ola_ar_fy25}, and disclosed a suspected employee fraud of over Rs 1 crore at the same subsidiary \citep{ola_ar_fy25}. Booking, invoice and registration are three different records of one month's truth, with a vendor dependency sitting inside the record chain. \caseinsight{When the definitions and the custody of the record are not architected, the market, the regulator and the company can each hold a different number in good faith. The failure is at Record and Retain.}

\textbf{Case D: Japan, one paragraph.} The pattern is neither Indian nor electric. On 3 June 2024 Toyota apologised for certification irregularities affecting seven models and suspended production of three, the Corolla Fielder, Corolla Axio and Yaris Cross, over faulty pedestrian-safety test data \citep{cnbc_toyota2024,fortune_toyota2024}; the transport ministry named five makers that day, Toyota, Mazda, Honda, Suzuki and Yamaha \citep{cnn_toyota2024}. In one test Toyota had used development data with a 65-degree impact angle where the regulation stipulated 50, which the company called a stricter condition and the ministry noted the angle alone does not determine \citep{asianews_toyota2024}. The chairman's own diagnosis was system overload; he said "We are not a perfect company" and attributed problems to standards stricter than certification required \citep{cnbc_toyota2024}. The case followed a December 2023 finding of 174 irregularities across 64 Daihatsu models dating to 1989 \citep{thedrive_daihatsu2023}, and earlier cases at Hino and Toyota Industries \citep{toyota_industries2024}. \caseinsight{The engineering held; the evidence procedure did not, at Create/Generate and Approve/Sign.} The companion paper develops this cluster into a full taxonomy.

%% file: tables/tab3_cases.tex
\begin{table*}[t]
\centering
\footnotesize
\setlength{\tabcolsep}{4pt}
\renewcommand{\arraystretch}{1.2}
\caption{The four failure cases mapped to the lifecycle stage (\cref{fig:lifecycle}) at which the evidence chain broke, the failure type, and the observed consequence. Contested matters are stated as contested; each company's position appears in the Section~5 text.}
\label{tab:cases}
\begin{tabularx}{\textwidth}{@{}L{2.9cm} L{1.5cm} L{2.5cm} L{3.2cm} X@{}}
\toprule
\headrow
\textbf{Case} & \textbf{Year(s)} & \textbf{Lifecycle stage} & \textbf{Failure type} & \textbf{Observed consequence} \\
\midrule
A. India e-scooter fires and recalls & 2022 & Retrieve / Reconstruct (locus A) & Post-incident evidence reconstruction across systems & 6,656 vehicles recalled across three makers; cell-batch to test-report to BMS-version to approver chain reconstructed under load \\
B. FAME-II localisation dispute & 2022--2025 & Verify / Reconcile (locus B) & State agencies reaching conflicting conclusions from one evidence base & Subsidy halted, notices, an April 2024 clean chit, debarments and a reported SFIO investigation coexisting over three years \\
C. Ola February 2025 registration gap & 2025 & Record / Retain (locus C) & Booking, invoice and registration as three competing records, with a vendor dependency in the record chain & About 25,000 announced against roughly 8,600 VAHAN registrations; government asked for an invoiced-basis restatement \\
D. Japan certification irregularities & 2024 & Create/Generate and Approve/Sign (locus D) & Certification test-evidence procedure, not the engineering, non-conformant & Production of models suspended across several makers after irregular certification data \\
\bottomrule
\end{tabularx}
\end{table*}

%% file: figures/fig3_timeline.tikz
\begin{tikzpicture}[
  font=\scriptsize,
  ev/.style={rectangle, rounded corners=1.5pt, draw=sapphireDark, fill=sapphireLight,
             line width=0.5pt, inner sep=2pt, text width=2.05cm, align=center},
  reg/.style={rectangle, rounded corners=1.5pt, draw=black!55, fill=black!5,
              line width=0.5pt, inner sep=2pt, text width=2.05cm, align=center},
  fame/.style={rectangle, rounded corners=1.5pt, draw=amber!80!black, fill=amber!18,
               line width=0.6pt, inner sep=2pt, text width=2.05cm, align=center},
  stem/.style={draw=sapphireMid, line width=0.5pt},
  rstem/.style={draw=black!55, line width=0.5pt},
  fstem/.style={draw=amber!80!black, line width=0.6pt},
]
\draw[-{Stealth[length=2mm]}, line width=1pt] (-0.3,0) -- (16.6,0);
\foreach \x/\yr in {0.6/2022, 3.6/2023, 6.6/2024, 9.6/2025, 12.6/2026, 15.4/2027--28}{
  \draw[line width=0.7pt] (\x,0.12) -- (\x,-0.12);
  \node[below, font=\scriptsize\bfseries] at (\x,-0.16) {\yr};
}
\node[right, font=\scriptsize, text=black!60] at (15.7,0.28) {(future)};

\node[ev]   (e1) at (0.9,1.5)  {e-scooter fires \& recalls (A)};   \draw[stem] (0.9,0)  -- (e1.south);
\node[fame] (e2) at (2.9,2.6)  {FAME-II probe, subsidy halt (B)}; \draw[fstem] (2.4,0) -- (e2.south);
\node[fame] (e3) at (5.6,2.6)  {clean chit vs.\ debarment (B)};   \draw[fstem] (6.2,0) -- (e3.south);
\node[ev]   (e4) at (7.9,1.5)  {Toyota certification (D)};        \draw[stem] (6.9,0)  -- (e4.south);
\node[fame] (e5) at (8.3,3.5)  {SFIO raids (B)};                  \draw[fstem] (7.4,0) -- (e5.south);
\node[ev]   (e6) at (10.4,1.5) {Ola registration gap (C)};        \draw[stem] (9.9,0)  -- (e6.south);
\draw[amber!80!black, line width=0.8pt, densely dashed] (e2.east) .. controls +(0.8,0) and +(-0.8,0) .. (e3.west);
\draw[amber!80!black, line width=0.8pt, densely dashed] (e3.north) .. controls +(0,0.4) and +(-0.6,0) .. (e5.west);

\node[reg] (r1) at (1.6,-1.5)  {AIS-156/038 amendments phased};  \draw[rstem] (1.4,0) -- (r1.north);
\node[reg] (r2) at (4.2,-1.5)  {AIS-156 Amd 4};                  \draw[rstem] (4.6,0) -- (r2.north);
\node[reg] (r3) at (7.0,-2.6)  {PM E-DRIVE scheme};              \draw[rstem] (7.0,0) -- (r3.north);
\node[reg] (r4) at (9.4,-1.5)  {ADAS GSR 184(E) draft};          \draw[rstem] (9.8,0) -- (r4.north);
\node[reg] (r5) at (11.6,-2.6) {Battery Pack Aadhaar draft};     \draw[rstem] (11.4,0) -- (r5.north);
\node[reg, draw=black!45, densely dashed] (r6) at (15.2,-1.6) {CSMS/SUMS/ADAS phase-in};
  \draw[rstem, densely dashed] (15.2,0) -- (r6.north);

\node[anchor=west, text=sapphireDark, font=\scriptsize\bfseries] at (-0.3,3.9) {Failure events};
\node[anchor=west, font=\scriptsize\bfseries] at (-0.3,-3.3) {Regulatory response};
\end{tikzpicture}

%% file: figures/fig1_lifecycle.tikz
\begin{tikzpicture}[
  font=\small,
  stage/.style={rectangle, rounded corners=2pt, draw=sapphireDark, line width=0.7pt,
                fill=sapphireLight, minimum height=1.05cm, text width=2.0cm, align=center,
                inner sep=3pt},
  flow/.style={-{Stealth[length=2.2mm]}, draw=sapphireMid, line width=0.9pt},
  loop/.style={-{Stealth[length=2.2mm]}, draw=sapphireMid, line width=0.8pt, densely dashed},
  mbox/.style={circle, draw=sapphireDark, fill=white, line width=0.6pt, inner sep=0pt,
               minimum size=3.8mm, font=\scriptsize\bfseries, text=sapphireDark},
  pin/.style={rectangle, rounded corners=1pt, draw=amber!80!black, fill=amber!18,
              line width=0.6pt, inner sep=2.5pt, font=\scriptsize, text=black},
]

\node[stage] (create)   {Create /\\ Generate};
\node[stage, right=0.8cm of create]  (submit)   {Submit /\\ Exchange};
\node[stage, right=0.8cm of submit]  (verify)   {Verify /\\ Reconcile};
\node[stage, right=0.8cm of verify]  (approve)  {Approve /\\ Sign};
\node[stage, right=0.8cm of approve] (record)   {Record /\\ Retain};
\node[stage, right=0.8cm of record]  (retrieve) {Retrieve /\\ Reconstruct};

\draw[flow] (create)  -- (submit);
\draw[flow] (submit)  -- (verify);
\draw[flow] (verify)  -- (approve);
\draw[flow] (approve) -- (record);
\draw[flow] (record)  -- (retrieve);

\coordinate (cut) at ($(approve.east)!0.5!(record.west)$);
\draw[line width=0.9pt, black] ($(cut)+(-1.3mm,2.1mm)$) -- ($(cut)+(1.3mm,-2.1mm)$);
\draw[line width=0.9pt, black] ($(cut)+(0.7mm,2.1mm)$)  -- ($(cut)+(3.3mm,-2.1mm)$);
\node[font=\scriptsize, text width=2.1cm, align=center, below=1.2mm of cut]
  {approval decouples\\ from artefact};

\draw[loop] (retrieve.north) .. controls +(0,1.25cm) and +(0,1.25cm) .. (create.north)
  node[midway, above, font=\scriptsize, text=sapphireDark] {Change / Revalidate loop};

\node[mbox] at (verify.north east)   {M};
\node[mbox] at (approve.north east)  {M};
\node[mbox] at (retrieve.north east) {M};

\node[pin] (locD) at ($(create.south)+(0,-0.95cm)$)   {D: Japan test-evidence};
\node[pin] (locB) at ($(verify.south)+(0,-0.95cm)$)   {B: FAME-II reconcile};
\node[pin] (locC) at ($(record.south)+(0,-0.95cm)$)   {C: Ola record};
\node[pin] (locA) at ($(retrieve.south)+(0,-0.95cm)$) {A: 2022 reconstruct};
\draw[amber!80!black, line width=0.5pt] (create.south)   -- (locD.north);
\draw[amber!80!black, line width=0.5pt] (approve.south)  .. controls +(0,-0.5cm) and +(0,0.4cm) .. (locD.north east);
\draw[amber!80!black, line width=0.5pt] (verify.south)   -- (locB.north);
\draw[amber!80!black, line width=0.5pt] (record.south)   -- (locC.north);
\draw[amber!80!black, line width=0.5pt] (retrieve.south) -- (locA.north);

\node[font=\scriptsize, anchor=west] at ($(create.north west)+(-0.05cm,1.75cm)$)
  {\tikz\node[mbox]{M};\,manual today \quad
   \tikz\draw[loop] (0,0) -- (0.5,0);\,revalidation loop \quad
   \tikz\node[pin]{};\,failure locus (Section 5)};

\end{tikzpicture}

%% file: sections/06_lifecycle.tex
\section{A lifecycle model of the compliance document}
\label{sec:lifecycle}

The four cases look like four different problems. The lifecycle model shows they are four instances of one. Every compliance artefact in this domain, a PPAP package, a battery test report, a software-update record, a conformity-of-production certificate, moves through the same six stages, shown in Figure~\ref{fig:lifecycle}: Create and Generate, Submit and Exchange, Verify and Reconcile, Approve and Sign, Record and Retain, and Retrieve and Reconstruct, with a Change and Revalidate loop that returns a changed artefact to the start. The model is deliberately artefact-centric rather than process-centric: it follows the document, not the business process, because the document is what the obligations attach to and what the failures happen to.

Three features of the model carry the argument. First, several stages are still manual today. Verify and Reconcile is manual because it means a human reading one artefact against others, a test report against the type it certifies, a PSW against a drawing revision \citep{aiagppap}, and checking objective but scattered fields \citep{ppapreject}. Approve and Sign is manual because it is a human judgement. Retrieve and Reconstruct is manual, and worst of all, because it is performed rarely, under pressure, and against records that were filed by people who have often moved on. Figure~\ref{fig:lifecycle} marks these bottlenecks.

Second, approval decouples from the artefact. The signature, the approval record, is frequently a separate object from the thing it approves: a warrant disposition in one system, a certificate in another, an email in a third, while the test report sits in a fourth. The binding between them is by reference, by a part number, a revision, a date, and references drift. This decoupling, marked in Figure~\ref{fig:lifecycle}, is why the reconstruction question in Case A is hard: the approver and the approved must be re-joined after the fact.

Third, the failure loci sit at specific stages. Case A is a Retrieve and Reconstruct failure (locus A): the artefacts existed but could not be rejoined on demand. Case B is a Verify and Reconcile failure (locus B): the same evidence base was reconciled to opposite conclusions by different agencies. Case C is a Record and Retain failure (locus C): three record systems, booking, invoice and registration, held three truths, with a vendor dependency inside the chain. Case D is a Create/Generate and Approve/Sign failure (locus D): the evidence was generated off-specification and approved anyway.

To make the model concrete, follow one artefact. A cell supplier runs cell-level abuse and safety tests and issues a battery test report (Create), then exchanges it with the pack assembler and the OEM (Submit). The OEM's homologation team reconciles it against the pack design, the applicable AIS revision, the state of charge tested, and the BMS configuration (Verify), an entirely manual reading-across. A signatory dispositions it, and that disposition is recorded apart from the report (Approve, and the decoupling). Both are filed and must stay retrievable for a horizon that, for the software-linked records alongside them, runs a decade past end of production (Record and Retain). If the cell is revised, the artefact re-enters at Create through the loop, and the new version must supersede the old without breaking the chain. When a field incident occurs, someone must pull that exact report, for that batch and firmware version, and prove the approval attached to it (Retrieve and Reconstruct). The four cases are what happens when one stage is overwhelmed, and they turn the anecdotes into the requirements the next section derives.

%% file: sections/07_lens.tex
\section{The exergy-destruction lens}
\label{sec:lens}

We propose an analytic lens for the work catalogued above. In thermodynamics, exergy is the maximum useful work obtainable from a system as it is brought into equilibrium with its environment \citep{bejan1997}. Every real, irreversible process destroys part of that available work, and because real processes are always irreversible the destruction is structural and strictly positive: it can be reduced by better process design but never driven to zero \citep{bejan1997}. The Gouy-Stodola theorem fixes its magnitude, equating the destroyed exergy to the dead-state temperature multiplied by the entropy generated \citep{bejan1997,moran2018}.

Regulated manufacturing has an analogous quantity. Alongside the useful work of designing, building and selling a compliant vehicle, the regulatory environment mandates an accompanying dissipation: the verification, reconciliation, approval and record-keeping effort that must be expended for the product to be legitimate, that scales with process complexity and regulatory density, and that never appears in the product itself. We call this documentation exergy destruction. The borrowing is explicitly analogical, an analogy and not an identity: we take the structure of the concept, a mandatory, structurally irreducible, minimisable-in-practice dissipation that accompanies useful work, and not its mathematics, its units, or any claim to a conserved quantity.

The lens classifies document families along four dimensions: verification effort per artefact; reconciliation complexity, meaning the number of other artefacts against which the artefact must be checked; retention horizon; and audit probability and consequence. High-destruction families score high on every axis. A Level 3 PPAP package must be checked against drawings, control plans and capability studies, and its approval gates shipment \citep{aiagppap}. A homologation evidence pack must reconcile a test report against the type it certifies. A software-update record set carries a retention obligation of at least ten years counted from when production of the vehicle type is definitely discontinued \citep{ais190}. Low-destruction families, such as a routine delivery note, score low on all four. Appendix B classifies the domain's families along these dimensions.

The lens does analytic work in two directions. First, it explains systematic under-resourcing. Work that is mandatory, output-adjacent and invisible in the finished product attracts neither talent nor tooling investment; it presents to management as a cost centre with no visible output, so it is starved, and starved verification is precisely the condition under which the failures of Section 5 occur. Second, it reframes those failures. They are not isolated lapses of diligence but the observable symptom of unmanaged destruction. When documentation exergy is generated faster than an organisation's manual processes can dissipate it in an orderly way, the backlog surfaces: an unreconstructable evidence chain (Case A), a reconciliation the state's own agencies cannot settle (Case B), competing records of truth with a vendor dependency in the record chain (Case C), or a certification procedure that breaks while the engineering holds (Case D).

Two caveats bound the analogy. It is qualitative: we do not assign the dissipation a temperature, an entropy, or a conserved balance, and no numeric exergy is computed. And it is a lens, not a law: its value is that it identifies where destruction concentrates and anticipates which document families are most exposed and why the failures recur, generating expectations that further cases can test, not that it quantifies them. That classificatory, hypothesis-generating role, made concrete in the Appendix B classification, is what distinguishes it from a metaphor.

%% file: sections/08_requirements.tex
\section{Evidence-architecture requirements}
\label{sec:requirements}

The lifecycle model and the cases yield a set of requirements on any evidence architecture that would manage this document domain well. We state them as neutral requirements, derived from the failures rather than from any product, and each is anchored to the lifecycle stage and the case that motivates it. They are necessary conditions, not a design.

\textbf{A1. Version anchoring.} Every artefact must be bound to the exact version of the thing it concerns: the drawing revision, the AIS amendment, the software version, the pack configuration. Motivated by Verify and Reconcile, where a wrong-revision mismatch is a leading rejection cause, and by Case A, where reconstruction turned on identifying the exact batch and firmware version.

\textbf{A2. Tamper-evidence.} An artefact and its history must be verifiable as unaltered since creation, so that a record retrieved years later can be trusted as what it purports to be. Motivated by Record and Retain and Retrieve and Reconstruct, and by the archival requirement that authenticity survive transmission and time.

\textbf{A3. Approval-to-artefact binding.} An approval must be inseparably bound to the specific artefact version it approves, rather than referenced across systems. Motivated directly by the approval-decoupling feature of the lifecycle model and by Case A, where the approver and the approved had to be rejoined after the fact.

\textbf{A4. Cross-system traceability.} The architecture must maintain coherent references across the MES, ERP, PLM and QMS systems in which evidence is generated and held \citep{qmsintegration}, so that an artefact can be followed across custody boundaries. Motivated by the distributed nature of the spine and by Case C, where booking, invoice and registration lived in separate systems that disagreed.

\textbf{A5. Long-horizon retention with integrity.} Records must remain retrievable and integrity-checkable for their full obligation horizon, which for software-update evidence extends to at least ten years past end of production. Motivated by Record and Retain, and by the software-traceability obligation of Section 4, whose horizon can outlast the systems that created the records.

\textbf{A6. Reliable human-in-the-loop verification.} Because Verify and Approve remain human judgements, the architecture must make the human step reliable at volume: surfacing the objective checks automatically, flagging the reconcilable mismatches, and reserving human attention for genuine judgement. Motivated by Verify and Reconcile as the dominant manual bottleneck, and by Case B, where reconciliation at scale failed.

\textbf{A7. Cross-party provenance.} The evidence chain spans a supplier, an OEM, a test agency and a regulator, each producing, holding and handing off distinct artefacts. The architecture must carry provenance across those parties, recording who produced, transformed and approved each artefact. Motivated by Case B, where cross-party localisation evidence could not be authoritatively settled, and by Case D, where a certification artefact crossed from producer to approver carrying an unflagged non-conformance.

These seven requirements are not independent wishes; they are the conditions under which the four failures would not have occurred, read off the stages at which they occurred. An architecture meeting A1 to A7 would let the Case A reconstruction be a query rather than an excavation, let the Case B reconciliation be settled once against anchored evidence, let the Case C records reconcile by construction, and let a Case D non-conformance be caught at the binding step rather than after production. The next section turns the gaps between this ideal and current practice into research problems.

%% file: sections/09_open_problems.tex
\section{Open problems and research agenda}
\label{sec:open}

The requirements of Section 8 expose problems that current research does not solve. For each we state why it is open and what evidence would count as progress.

\textbf{P1. Reconciliation accuracy, distinct from extraction accuracy.} A mature toolchain now extracts structure from documents: open toolkits such as Docling \citep{docling2024}, a landscape of parsers including Unstructured, Marker, LlamaParse and Reducto \citep{unstructured,marker,llamaparse,reducto}, cloud document-understanding services \citep{gcdocai,awstextract,azuredocai}, and vision-based retrieval in the ColPali line \citep{colpali2024}. But extraction is not the compliance task. The compliance task is reconciliation: does this test report satisfy this obligation for this version, checked against these other artefacts. Extraction accuracy and reconciliation accuracy are different quantities, and the field measures only the first. Progress would be a public benchmark for compliance reconciliation. None exists: DocVQA evaluates question answering over document images and RD-TableBench evaluates table extraction, and a targeted search found no benchmark that evaluates reconciliation of evidence against requirements \citep{docvqa2021,rdtablebench}. Constructing one, with real obligation-artefact pairs and reconciliation verdicts, is itself a research contribution.

\textbf{P2. Human-in-the-loop verification reliability under volume.} Because Verify and Approve remain human, the open question is how reliable a human verifier is across thousands of artefacts, how error rate scales with volume and fatigue, and how to allocate attention so that machine-checkable items are pre-cleared and human judgement is spent where it matters. There is no measurement model for this in the compliance-evidence setting; progress would be one, validated against observed verification error.

\textbf{P3. Cross-party approval provenance.} The chain crosses supplier, OEM, test agency and regulator, and Case B showed that when provenance is weak, the state's own agencies can disagree over years. The open problem is a provenance model that binds an approval to an artefact version across organisational boundaries, verifiable by a later party without trusting the intermediate ones. Progress would be a scheme that lets a regulator confirm, from the record alone, who approved what version on what evidence.

\textbf{P4. Long-horizon evidence integrity.} AIS-190, mirroring UN R156, requires software-update documentation to be retained and remain available for at least ten years after production of the type is discontinued \citep{ais190,ais189}. This is an unusual class of obligation: the integrity and retrievability of a record must outlast the systems, formats and sometimes the organisations that created it. The open problem is integrity over horizons longer than technology cycles; progress would be demonstrated retrievability and tamper-evidence of a software-provenance record across a format and system migration.

\textbf{P5. Measuring the documentation burden.} The burden this paper describes has no denominator. There is no published measure of how many artefacts an OEM produces per model, how much verification labour they consume, or what fraction of engineering effort the evidence layer absorbs. Without measurement, the burden cannot be managed or its reduction demonstrated. Progress would be an empirical burden metric and a baseline for even one OEM programme.

\textbf{P6. Machine-verifiable regulatory design.} Finally, a forward problem for regulators. The obligations of Section 4 are written in prose and verified by humans. Many are objective enough to be specified machine-verifiably, so that conformance could be checked by construction rather than by after-the-fact reading. Progress would be a machine-readable specification of one evidence obligation, and a demonstration that conformance can be checked automatically against real artefacts.

These six are one agenda, not six. They share a centre: the compliance-evidence artefact as a first-class object, with accuracy, provenance, integrity, burden and verifiability defined on it. That object is what the four literatures of Section 2 leave unclaimed, and what this systematisation exists to put on the table.

%% file: sections/10_conclusion.tex
\section{Conclusion}
\label{sec:conclusion}

India's electric-vehicle transition has produced a documentation and traceability burden that is real, heavy and growing, and that no one had systematised from an information-systems perspective. We have done so for the Indian EV OEM. We organised the evidence obligations into two layers, the inherited quality spine and the EV-acute regulatory stack, by artefact, producer, verifier, trigger and retention horizon. We abstracted four documented failures into a single lifecycle model of the compliance document, marking where the work is manual, where approval decouples from the artefact, and where each failure occurred. We proposed the exergy-destruction lens, an explicitly analogical framing that classifies document families by the mandatory, output-adjacent verification and record-keeping work they carry and explains why that work is systematically under-resourced. And we set out a research agenda whose common object is the compliance-evidence artefact itself: reconciliation accuracy as distinct from extraction, human-in-the-loop reliability at volume, cross-party approval provenance, evidence integrity over decade-long horizons, the missing measurement of burden, and machine-verifiable regulatory design. The manual verification bottleneck this domain runs on is not a detail of implementation; it is where the failures live. A companion paper extends the quality spine and a global failure-mode taxonomy. The burden is now mapped; it can be built on.

%% file: sections/appendix_a_standards.tex
\section{Standards inventory at the snapshot}
\label{app:standards}
Table~\ref{tab:inventory} inventories the instruments referenced in the paper, with scope,
the key evidence obligations each imposes, and status as of the snapshot date. Every row is
traceable to the claims ledger and is cited in \cref{sec:spine,sec:evstack}.

\begin{table}[h]
\centering
\footnotesize
\setlength{\tabcolsep}{4pt}
\renewcommand{\arraystretch}{1.2}
\caption{Full standards inventory. Dates and status stated as of \snapshotdate.}
\label{tab:inventory}
\begin{tabularx}{\textwidth}{@{}L{2.7cm} L{2.2cm} X L{3.4cm}@{}}
\toprule
\headrow
\textbf{Instrument} & \textbf{Scope} & \textbf{Key evidence obligations} & \textbf{Dates / status} \\
\midrule
IATF 16949:2016 \citep{iatf16949} & Automotive suppliers & Documented-information control, defined retention policy, identification and traceability & In force \\
AIAG PPAP (4th ed.) \citep{aiagppap} & Part approval & Up to 18 elements, five submission levels, PSW gating shipment & In force \\
EN 10204 \citep{en10204} & Material inspection & Types 3.1 (manufacturer) and 3.2 (independent) inspection documents & In force \\
AIS-038 (Rev.2) \citep{ais038} & M, N EVs & Traction-battery safety, thermal-propagation and IPX7 evidence, pack/cell/BMS/charger traceability & Phased mandatory 1 Dec 2022 and 31 Mar 2023 \citep{morth_amend2022} \\
AIS-156 \citep{ais156} & L EVs & REESS and power-train safety, audio-visual thermal warning, smart BMS, temperature sensors & As above; Amendment 4 (Dec 2023) \\
CMVR type approval / CoP \citep{cmvr_gsr659e} & All categories & Type approval (Rule 126, AIS-049 Rev.1); conformity of production for traction batteries & CoP effective 1 Oct 2022 \\
AIS-189 (CSMS $\sim$ UN R155) \citep{ais189} & Road vehicles & Cyber-security management-system process and risk-assessment evidence & Phased schedule proposed; gazette dates not settled \\
AIS-190 (SUMS $\sim$ UN R156) \citep{ais190} & Road vehicles & Software-update management-system package; version/update records; retention $\geq$10 years post end-of-production & Proposed; gazette dates not settled \\
ADAS suite AIS-162/184/186/187/188 \citep{morth_gsr184e} & M2, M3, N2, N3 & Function type-approval evidence (AEBS, drowsiness, blind-spot, moving-off, lane-departure) & Draft GSR 184(E) (20 Mar 2025); phase-in 2027--2028 \\
Battery Pack Aadhaar (draft) \citep{bpan_draft} & L, M, N and $>$2 kWh industrial & Unique battery identity and lifecycle traceability data & Draft (OM 17 Sep 2025); reported 2027 \\
Battery Waste Management Rules 2022 \citep{bwmr2022} & Battery producers & Extended producer responsibility; CPCB registration; collection/recycling records & In force \\
EU Regulation 2023/1542 (comparator) \citep{eu2023_1542} & EV, LMT, $>$2 kWh industrial & Battery passport (Article 77) & Applies from 18 Feb 2027 \\
\bottomrule
\end{tabularx}
\end{table}

%% file: sections/appendix_b_taxonomy.tex
\section{Document-family taxonomy under the lens}
\label{app:taxonomy}
Table~\ref{tab:taxonomy} applies the exergy-destruction lens of \cref{sec:lens} to the domain's
document families, scoring each along the four lens dimensions and assigning a destruction class.
High-destruction families score high on verification effort, reconciliation complexity, retention
horizon, and audit probability or consequence; low-destruction families score low on all four.
The classification is the concrete, testable payload of the lens: it predicts which families
absorb the most mandatory verification and record-keeping work, and therefore which fail first
when that work is under-resourced.

\begin{table}[h]
\centering
\footnotesize
\setlength{\tabcolsep}{4pt}
\renewcommand{\arraystretch}{1.2}
\caption{Document families classified along the lens dimensions (verification effort; reconciliation complexity; retention horizon; audit probability/consequence) with a destruction class.}
\label{tab:taxonomy}
\begin{tabularx}{\textwidth}{@{}X L{1.7cm} L{2.3cm} L{2.5cm} L{1.9cm} L{1.4cm}@{}}
\toprule
\headrow
\textbf{Document family} & \textbf{Verif.\ effort} & \textbf{Reconcil.\ complexity} & \textbf{Retention horizon} & \textbf{Audit prob./cons.} & \textbf{Class} \\
\midrule
PPAP Level 3 package & High & High (drawings, control plans, capability) & Programme life & High & High \\
PSW / approval disposition & Medium & Medium & Programme life & High & High \\
Homologation / type-approval evidence pack & High & High (report vs type) & Type-approval validity & High & High \\
Battery thermal-propagation test report & High & Medium & Long & High & High \\
Software-update / version record set & Medium & High (version chain) & $\geq$10 yr post-production & High & High \\
Conformity-of-production records & Medium & High (serial vs prototype) & Ongoing production & High & High \\
Material certificate 3.1 / 3.2 & Low--Med & Medium & Order / product life & Medium & Medium \\
EPR / battery-passport records & Medium & Medium & Battery life & Medium & Medium \\
Calibration record & Low & Low & Periodic cycle & Medium & Low--Med \\
Routine delivery note & Low & Low & Short & Low & Low \\
\bottomrule
\end{tabularx}
\end{table}

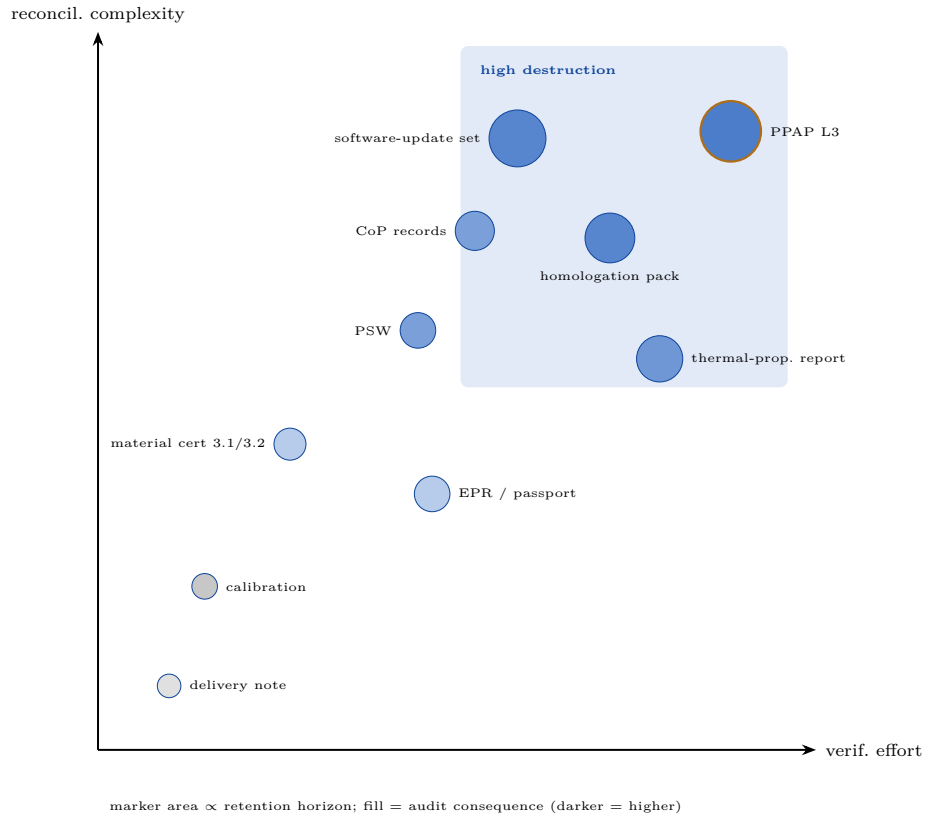
\begin{figure}[h]
\centering
\resizebox{0.72\linewidth}{!}{\input{figures/fig4_lens.tikz}}
\caption{Document families under the exergy-destruction lens, plotting the classification of \cref{tab:taxonomy}. Position gives verification effort and reconciliation complexity; marker area gives retention horizon; colour gives audit consequence. High-destruction families cluster top right.}
\label{fig:lens}
\end{figure}

%% file: figures/fig4_lens.tikz
\begin{tikzpicture}[
  font=\scriptsize,
  fam/.style={circle, draw=sapphireDark, line width=0.4pt},
  lbl/.style={font=\tiny, text=black, inner sep=1.5pt},
]
\begin{scope}[on background layer]
  \fill[sapphireLight, rounded corners=3pt] (5.1,5.1) rectangle (9.7,9.9);
\end{scope}
\node[font=\tiny\bfseries, text=sapphireDark, anchor=north west] at (5.25,9.75) {high destruction};

\draw[-{Stealth[length=2mm]}, line width=0.8pt] (0,0) -- (10.1,0)
  node[right, font=\scriptsize] {verif.\ effort};
\draw[-{Stealth[length=2mm]}, line width=0.8pt] (0,0) -- (0,10.1)
  node[above, font=\scriptsize] {reconcil.\ complexity};

\newcommand{\famnode}[6]{\node[fam, fill=#4, minimum size=#3] (n) at (#1,#2){}; \node[lbl, #6=1.5pt of n]{#5};}

\famnode{1.0}{0.9}{3mm}{black!12}{delivery note}{right}
\famnode{1.5}{2.3}{3.6mm}{black!22}{calibration}{right}
\famnode{2.7}{4.3}{4.5mm}{sapphire!30}{material cert 3.1/3.2}{left}
\famnode{4.7}{3.6}{5mm}{sapphire!30}{EPR / passport}{right}
\famnode{4.5}{5.9}{5mm}{sapphire!55}{PSW}{left}
\famnode{5.3}{7.3}{5.5mm}{sapphire!55}{CoP records}{left}
\famnode{7.9}{5.5}{6.5mm}{sapphire!60}{thermal-prop.\ report}{right}
\famnode{5.9}{8.6}{8mm}{sapphire!70}{software-update set}{left}
\famnode{7.2}{7.2}{7mm}{sapphire!70}{homologation pack}{below}
\node[fam, draw=amber!80!black, line width=0.9pt, fill=sapphire!75, minimum size=8.5mm] (ppap) at (8.9,8.7){};
\node[lbl, right=1.5pt of ppap]{PPAP L3};

\node[lbl, anchor=west] at (0.1,-0.8) {marker area $\propto$ retention horizon;\ fill $=$ audit consequence (darker $=$ higher)};
\end{tikzpicture}

%% file: sections/appendix_c_chronologies.tex
\section{Case chronologies and sources}
\label{app:chronologies}
The chronologies below expand \cref{sec:failures}. Contested matters remain stated as contested;
reported investigations are described as reported. Sources are the citations shown.

\paragraph{Case A: 2022 fires and recalls.}
\begin{itemize}
\item 16 Apr 2022: Okinawa Autotech recalls 3,215 Praise Pro scooters \citep{recalls2022}.
\item 21 Apr 2022: Pure EV recalls 2,000 scooters (ETrance+, EPluto 7G) \citep{recalls2022}.
\item 23 Apr 2022: Ola Electric recalls 1,441 scooters after the Pune S1 Pro fire; government total 6,656 \citep{recalls2022}.
\item 2022: show-cause notices to makers; expert committee (DRDO/CFEES, NSTL, IISc, ARCI, IIT-Madras) convened \citep{evfires_committee2022}.
\item From 1 Dec 2022 and 31 Mar 2023: AIS-156 and AIS-038 (Rev.2) amendments phased in as the response \citep{morth_amend2022}.
\end{itemize}

\paragraph{Case B: FAME-II localisation dispute.}
\begin{itemize}
\item 2019 onward: FAME-II scheme, outlay about Rs 10,000 crore (later enhanced) \citep{bt_recovery2023}.
\item 2022--2023: probe of around a dozen e2W makers over alleged localisation shortfalls; subsidy on hold for several \citep{bt_recovery2023,bs_famedenial2023}.
\item May 2023: show-cause notices reported to six makers seeking about Rs 469 crore; three later repaid with interest \citep{bt_recovery2023}.
\item 27 Mar 2024: debarment orders reported against Hero Electric and Benling; deregistration from FAME-II \citep{inc42_cleanchit}.
\item Apr 2024: ministry committee clean chit to Hero Electric and Okinawa, citing ARAI (13 May 2022) and ARAI+ICAT (8 Aug 2022) reports and guideline ambiguity \citep{bt_cleanchit2024,inc42_cleanchit}.
\item From Dec 2024: SFIO search operations reported at the three firms over about Rs 297 crore \citep{sfio_probe}.
\item 31 Mar 2024: FAME-II ends; succeeded by EMPS 2024 and PM E-DRIVE \citep{pmedrive}.
\end{itemize}

\paragraph{Case C: Ola February 2025 registration gap.}
\begin{itemize}
\item 19 Feb 2025: Ola exchange filing on registration-vendor renegotiation (Rosmerta, Shimnit) and temporary VAHAN impact; figure framed as bookings \citep{ola_filing2025}.
\item Feb 2025: about 25,000 (25,207) announced against roughly 8,600 VAHAN registrations \citep{bs_ola_feb2025}.
\item Reported: bookings included about 10,866 Gen-3 scooters and 1,395 Roadster X; 31 Mar 2025 government letter seeking invoiced-basis restatement \citep{outlook_ola2025}.
\item Mar--Apr 2025: RTO trade-certificate seizures reported (36 in Mumbai and Pune, escalating to about 192) \citep{bs_ola_feb2025}.
\item FY25 annual report: statutory-auditor material weakness in physical-verification controls at a subsidiary (about Rs 362 crore); suspected employee fraud over Rs 1 crore disclosed \citep{ola_ar_fy25}.
\end{itemize}

\paragraph{Case D: Japan certification irregularities.}
\begin{itemize}
\item 29 Jan 2024: Toyota Industries diesel-engine certification irregularities; 10 models suspended \citep{toyota_industries2024}.
\item 3 Jun 2024: Toyota apologises for irregularities across seven models; suspends Corolla Fielder, Corolla Axio, Yaris Cross \citep{cnbc_toyota2024,fortune_toyota2024}.
\item 3 Jun 2024: transport ministry names five makers (Toyota, Mazda, Honda, Suzuki, Yamaha) \citep{cnn_toyota2024}.
\item Pedestrian test: 65-degree impact angle used where regulation stipulated 50; ministry caveat on strictness \citep{asianews_toyota2024}.
\item Context: December 2023 Daihatsu finding of 174 irregularities across 64 models dating to 1989 \citep{thedrive_daihatsu2023}.
\end{itemize}

%% file: sections/appendix_d_littable.tex
\section{Nearest-neighbour literature}
\label{app:littable}
Table~\ref{tab:litmap} lists the works surveyed for \cref{sec:background}, grouped by literature
(L1 regulatory, certification and transport policy; L2 battery safety and automotive engineering;
L3 information systems; L4 records and archival management), with what each misses relative to this
paper's target: an information-systems systematisation of the automotive and EV compliance-document
lifecycle, its evidence artefacts, obligations and failures. No surveyed work occupies that centre.

\footnotesize
\setlength{\LTleft}{0pt}\setlength{\LTright}{0pt}
\renewcommand{\arraystretch}{1.15}
\begin{longtable}{@{}L{4.2cm} L{0.7cm} L{11.2cm}@{}}
\caption{Nearest-neighbour survey (37 works). What each misses relative to this paper.}
\label{tab:litmap}\\
\toprule
\headrow
\textbf{Work} & \textbf{Lit} & \textbf{What it misses} \\
\midrule
\endfirsthead
\multicolumn{3}{l}{\emph{Table~\ref{tab:litmap} continued}}\\
\toprule
\headrow
\textbf{Work} & \textbf{Lit} & \textbf{What it misses} \\
\midrule
\endhead
\bottomrule
\endfoot
Sadiq, Governatori \& Namiri \citep{sadiq2007} & L3 & Models control objectives at design time; no evidence-document lifecycle or automotive/EV domain. \\
Governatori \& Sadiq \citep{governatori2009} & L3 & Process-versus-norm logic only; silent on the physical evidence documents. \\
Hashmi et al.\ \citep{hashmi2018} & L3 & Surveys process-centric compliance; the evidence artefact is never the unit of analysis. \\
Arner, Barberis \& Buckley \citep{arner2017} & L3 & RegTech vision for finance; no product-safety or automotive evidence. \\
Becker, Merz \& Buchkremer \citep{becker2020} & L3 & RegTech review, financial; no compliance-document artefact chain. \\
Xu et al.\ (LayoutLM) \citep{xu2020} & L3 & Form field extraction; agnostic to compliance semantics and provenance. \\
Appalaraju et al.\ (DocFormer) \citep{appalaraju2021} & L3 & Visual document understanding; no compliance modelling. \\
Cui et al.\ (Document AI) \citep{cui2021} & L3 & Technology survey; no link to obligations, records or domain. \\
Costantino et al.\ \citep{costantino2022} & L1 & Aligns R155 and ISO/SAE 21434; no document-evidence lifecycle. \\
Oberti et al.\ \citep{oberti2024} & L1 & Automotive-cybersecurity compliance overview; no records lifecycle. \\
Benyahya et al.\ \citep{benyahya2023} & L1 & CAV certification roadmap; no evidence-artefact lifecycle. \\
Seo, Kwak \& Kim \citep{seo2024} & L1 & Secure SUMS architecture; not the compliance-evidence artefacts. \\
Hellstern et al.\ \citep{hellstern2024} & L1 & Derives R155 approval criteria; not the document lifecycle. \\
Henle et al.\ \citep{henle2025} & L1 & Derives R156 compliance evidence; single regulation, engineering process, not an IS systematisation. \\
Yadav \& Sircar \citep{yadav2022} & L1 & India EV policy review; no certification-evidence documents. \\
Singh et al.\ \citep{singh2021} & L1 & EV adoption factors; no compliance-document lifecycle. \\
Vashist \& Pandey \citep{vashist2024} & L2 & Indian battery-safety standards and tests; documentation unexamined. \\
Ruiz et al.\ \citep{ruiz2018} & L2 & Abuse-testing standards comparison; test reports an unexamined by-product. \\
Hannan et al.\ \citep{hannan2017} & L2 & BMS and state-of-charge estimation; documentation not the object. \\
Mallick \& Gayen \citep{mallick2023} & L2 & Thermal-runaway physics; nothing on evidence artefacts. \\
Jaguemont \& Bard\'e \citep{jaguemont2023} & L2 & Safety-testing methods review; documentation is background. \\
Naiek et al.\ \citep{naiek2025} & L2 & BEV safety and policy; no compliance documents or failures. \\
Birch et al.\ \citep{birch2013} & L2 & ISO 26262 safety case as an assurance argument, not a records lifecycle. \\
Berger et al.\ \citep{berger2022} & L3 & Battery-passport data model; sustainability data, not compliance evidence. \\
Pohlmann et al.\ \citep{pohlmann2025} & L3 & DPP information requirements; sustainability-oriented. \\
Losa \& Torjesen \citep{losa2025} & L1 & EU battery-passport implementation; single instrument, policy lens. \\
Gie\ss{} \& M\"oller \citep{giess2025} & L3 & DPP ecosystem actors and value; no compliance-document lifecycle. \\
Lopes \& Barata \citep{lopes2024} & L3 & DPP review and agenda; object is DPPs generally, not compliance evidence. \\
Shen et al.\ \citep{shen2024} & L3 & Blockchain parts-traceability mechanism; not an artefact/obligation systematisation. \\
Lixandru \citep{lixandru2016} & L3 & Supplier-quality process description; not an IS evidence-lifecycle systematisation. \\
Farahani et al.\ \citep{farahani2016} & L3 & Digital-SCM agenda; does not treat compliance documentation. \\
ISO 15489-1:2016 \citep{iso15489} & L4 & General records-management principles; deliberately domain-agnostic. \\
Duranti \citep{duranti1995} & L4 & Reliability and authenticity concepts; no concrete artefact set. \\
Duranti (InterPARES) \citep{duranti2005} & L4 & Digital-authenticity framework; not this domain's artefacts or failures. \\
Upward \citep{upward1996} & L4 & Records continuum model; no domain instantiation. \\
McKemmish \citep{mckemmish2001} & L4 & Records continuum in practice; abstract, no domain map. \\
Hamilton \citep{hamilton1991} & L4 & Records management in engineering firms; predates digital compliance regimes. \\
\end{longtable}
\normalsize